# Data Mining on Educational Domain

*Nikhil Rajadhyax*
*Department of Information Technology*
Shree Rayeshwar Institute of Engineering & Information Technology
Shiroda, Goa, India
e-mail: nikhil.rajadhyax@gmail.com

*Prof. Rudresh Shirwaikar*
*Department of Information Technology*
Shree Rayeshwar Institute of Engineering & Information Technology
Shiroda, Goa, India
e-mail: rudreshshirwaikar@gmail.com

*Abstract*— **Educational data mining (EDM) is defined as the area of scientific inquiry centered around the development of methods for making discoveries within the unique kinds of data that come from educational settings , and using those methods to better understand students and the settings which they learn in.**
**Data mining enables organizations to use their current reporting capabilities to uncover and understand hidden patterns in vast databases. As a result of this insight, institutions are able to allocate resources and staff more effectively.**
**In this paper, we present a real-world experiment conducted in Shree Rayeshwar Institute of Engineering and Information Technology (SRIEIT) in Goa, India. Here we found the relevant subjects in an undergraduate syllabus and the strength of their relationship. We have also focused on classification of students into different categories such as good, average, poor depending on their marks scored by them by obtaining a decision tree which will predict the performance of the students and accordingly help the weaker section of students to improve in their academics.**
**We have also found clusters of students for helping in analyzing student's performance and also improvising the subject teaching in that particular subject.**

## I. INTRODUCTION

The advent of information technology in various fields has lead the large volumes of data storage in various formats like records, files, documents, images, sound, videos, scientific data and many new data formats. The data collected from different applications require proper method of extracting knowledge from large repositories for better decision making. Knowledge discovery in databases (KDD), often called data mining, aims at the discovery of useful information from large collections of data [1]. The main functions of data mining are applying various methods and algorithms in order to discover and extract patterns of stored data.

There are increasing research interests in using data mining in education. This new emerging field, called Data Mining on Educational Domain, concerns with developing methods that discover knowledge from data originating from educational environments. Educational Data Mining uses many techniques such as Decision Trees, Neural Networks, Naïve Bayes, K-Nearest neighbor, and many others.

Using these techniques many kinds of knowledge can be discovered such as association rules, classifications and clustering. The discovered knowledge can be used for prediction regarding the overall performance of the student. The main objective of this paper is to use data mining methodologies to study student's performance in their academics.

## II. METHODOLOGY

### A. Background

SRIEIT's undergraduate degree programme - B.E. - consists of three fields of specialization (i.e. Information Technology, Electronics and Telecommunication, and Computer Engineering). Each field of specialization offers students many subjects during eight different semesters within a period of four years. A student belongs to a batch and the batch is offered a number of subjects.

The performance of students in the different courses offered provides a measure of the students' ability to meet lecturer/institution's expectations. The overall marks obtained by students in the different subjects are utilized in our experiment in finding related subjects. The main objective is to determine the relationships that exist between different courses offered as this is required for optimizing the organization of courses in the syllabi. This problem is solved in two steps:
1. Identify the possible related subjects.
2. Determine the strength of their relationships and determine strongly related subjects.

In the first step, we utilized association rule mining [1] to identify possibly related two subject combinations in the syllabi which also reduces our search space. In the second step, we applied Pearson Correlation Coefficient [2] to determine the strength of the relationships of subject combinations identified in the first step.

Our experiment is based on the following hypothesis:
"Typically, a student obtains similar marks for related subjects".

This assumption is justified by the fact that similar marks obtained by a student for different subjects imply that the student is meeting the expectations of the course in a similar manner. This fact implies an inherent relationship between the courses.

For this experiment, we selected students of batches from 2009-2010 , in semesters 3-6 and 60 student in the three fields of specialization. The first step, finding possible related subjects, requires considering 2-subject combinations.
To do this we applied Association Rules Mining [1].Association Rule Mining and its application are discussed in sections B and C.

### B. Association Rule Mining

Firstly, 2-subject combinations were obtained using Apriori algorithm by using database of the form TABLE II. Then

Association rules were applied to the output of Apriori algorithm

Association rules are an important class of regularities that exists in databases. The classic application of association rules is the market basket analysis [1]. It analyzes how items purchased by customers are associated.

Table I illustrates a sales database with items purchases in each transaction. An example association rule is as follows:

pen => ink [support 75%, confidence 80%]

This rule says that 75% of customers buy pen and ink together, and those who buy pen buys ink 80% of the time.

TABLE I. INSTANCE OF A SALES DATABASE

| Transaction ID | Items |
|---|---|
| 111 | pen, ink, milk |
| 112 | pen, ink |
| 113 | pen, ink, juice |
| 114 | pen, milk |

Formally, the association rule-mining model can be stated as follows. Let I = {$i_1, i_2,...,i_m$} be a set of items. Let D be a set of transactions (the database), where each transaction d is a set of items such that d ⊂ I. An association rule is an implication of the form, X->Y, where X ⊂ I, Y ⊂ I, and X ∩ Y = 0. The rule X -> Y holds in the transaction set D with confidence c if c% of transactions which contain X in D also contains Y. The rule has support s in D if s% of the transactions in D contains X∪Y.

Given a set of transactions D (the database), the problem of mining association rules is to discover all association rules that have support and confidence greater than or equal to the user specified minimum support (called *minsup*) and minimum confidence (called *minconf*).

*C. Application of Association Rule Mining*

At SRIEIT, a student earns either a "pass" grade (that is, a student meets the minimum requirements for successful completion of the subject) or "failure" grade (that is, a student fails to meet the minimum requirements for successful completion of the subject) for every subject the student followed. A transaction table is considered consisting of students with their passed subjects (see TABLE II).

Our goal is to find the relationship between two subjects (i.e. $subject_i$ and $subject_j$ where i ≠ j) using association rule mining. That is, find association rules with the following format meeting a certain selection criteria.

$subject_i$ -> $subject_j$, where i ≠ j

TABLE II. DATABASE INSTANCE OF STUDENT

| Student | Student Passed Subjects |
|---|---|
| S1 | $subject_1$, $subject_2$, $subject_3$ |
| S2 | $subject_1$, $subject_2$, $subject_4$ |
| S3 | $subject_1$, $subject_5$ |

In creating the database, we considered only passed subjects due to the fact that no subject had a 100% failure. To identify all possible related subjects (not necessarily subjects with high pass rates), we ignored the support and considered only confidence measure. The confidence was sufficient to determine the possible related subjects (for instance, in the above rule, confidence provides us with the percentage of students that had passed $subject_i$ also passed $subject_j$).

We considered the average pass rate as the minimum confidence:

$$minconf = \frac{\sum(pass\_rate\_for\_each\_subject)}{number\_of\_subjects}$$

*D. Pearson Correlation Coefficient*

The Pearson Correlation Coefficient (r) measures the strength of the linear relationship between two continuous variables. We computed r and selected a threshold value (i.e. γ) to determine strong relationships.

The Pearson Correlation Coefficient (r) is computed as follows:

$$r = \frac{\sum_{i=1}^{n}(X_i - \bar{X})(Y_i - \bar{Y})}{(n-1)S_x S_y}$$

where:
- X, Y are two continuous variables,
- Sx and Sy are the standard deviations of X and Y, and
- $\bar{X}$ and $\bar{Y}$ are the mean values of X and Y.

The value of r is such that -1 < r < +1. The + and – signs are used for positive linear correlations and negative linear correlations, respectively. If there is no linear correlation or a weak linear correlation, r is close to 0. A correlation greater than 0.5 is generally described as strong, whereas a correlation less than 0.5 is generally described as weak.

*E. Application of Pearson Correlation Coefficient*

After experimentation we selected 0.5 for the threshold value (i.e. γ = 0.5) as a suitable estimate for determining a strong relationship. A subject combination (say $subject_i$ and $subject_j$ where i ≠ j) may contain a strong relationship (that is r ≥ 0.5 for $subject_i$, $subject_j$ permutation).

*F. Classification Algorithm*

Here we make use of decision tree to classify the data and the tree is obtained by making use of ID3 algorithm. A decision tree is a tree in which each branch node represents a choice between a number of alternatives, and each leaf node represents a decision. Decision tree starts with a root node on which it is for users to take actions. From this node, users split each node recursively according to decision tree learning algorithm. The final result is a decision tree in which each branch represents a possible scenario of decision and its outcome. We provide the collected data to the algorithm to create a model called as classifier. Once the classifier is built we can make use of it and can easily classify any student and can predict its performance.

ID3 is a simple decision tree learning algorithm. The basic idea of ID3 algorithm is to construct the decision tree by employing a top-down, greedy search through the given sets to test each attribute at every tree node. In order to select the attribute that is most useful for classifying a given sets, we introduce a metric - information gain. To find an optimal way to classify a learning set we need some function which provides the most balanced splitting. The information gain metric is such a function. Given a data table that contains attributes and class of the attributes, we can measure homogeneity (or heterogeneity) of the table based on the classes. The index used to measure degree of impurity is Entropy.
The Entropy is calculated as follows:

$$E(S) = \sum_j - p_j \log_2 p_j$$

Splitting criteria used for splitting of nodes of the tree is Information gain. To determine the best attribute for a particular node in the tree we use the measure called Information Gain. The information gain, Gain (S, A) of an attribute A, relative to a collection of examples S, is defined as

$$Gain(S,A) = E(S) - \sum_v \frac{|S_v|}{|S|} E(S_v)$$

The ID3 algorithm is as follows:
- Create a root node for the tree
- If all examples are positive, Return the single-node tree Root, with label = +.
- If all examples are negative, Return the single-node tree Root, with label = -.
- If number of predicting attributes is empty, then Return the single node tree Root, with label = most common value of the target attribute in the examples.
- Otherwise Begin
   - A = The Attribute that best classifies examples.
   - Decision Tree attribute for Root = A.
   - For each possible value, vi, of A,
      - Add a new tree branch below Root, corresponding to the test A = vi.
      - Let Examples (vi) be the subset of examples that have the value vi for A
      - If Examples (vi) is empty
        - Then below this new branch add a leaf node with label = most common target value in the examples
      - Else below this new branch add the subtree ID3 (Examples (vi), Target_Attribute, Attributes – {A})
- End
-Return Root

Here we used attendance and marks of 60 students from 3 branches each. (See TABLE IV).

*G. Clustering Algorithm*
DBSCAN is a density-based spatial clustering algorithm. By density-based we mean that clusters are defined as connected regions where data points are dense. If density falls below a given threshold, data are regarded as noise.
DBSCAN requires three inputs:
1. The data source
2. A parameter, Minpts- which is the minimum number of points to define a cluster.
3. A distance parameter, Eps- a distance parameter- if there are atleast Minpts within Eps of a point is a core point in a cluster.

Core Object: Object with at least MinPts objects within a radius 'Eps-neighborhood'
Border Object: Object that on the border of a cluster
NEps(p): {q belongs to D | dist(p,q) <= Eps}
Directly Density-Reachable: A point p is directly density-reachable from a point q w.r.t Eps, MinPts if p belongs to NEps(q)
|NEps (q)| >= MinPts
Density-Reachable: A point p is density-reachable from a point q w.r.t Eps, MinPts if there is a chain of points p1, …, pn, p1 = q, pn = p such that pi+1 is directly density-reachable from pi
Density-Connected: A point p is density-connected to a point q w.r.t Eps, MinPts if there is a point o such that both, p and q are density-reachable from o w.r.t Eps and MinPts.
It starts with an arbitrary starting point that has not been visited. This point's ε-neighborhood is retrieved, and if it contains sufficiently many points, a cluster is started. Otherwise, the point is labeled as noise. Note that this point might later be found in a sufficiently sized ε-environment of a different point and hence be made part of a cluster.
If a point is found to be part of a cluster, its ε-neighborhood is also part of that cluster. Hence, all points that are found within the ε-neighborhood are added, as is their own ε-neighborhood. This process continues until the cluster is completely found. Then, a new unvisited point is retrieved and processed, leading to the discovery of a further cluster or noise.
Here too we used attendance and marks of 60 students from 3 branches each. (See TABLE III).

### III. RESULT

*A. Observations*

The output of association rule mining and later Pearson coefficient correlation provided us with the possibly related 2-subject combination and the strength of their relationship.
The subjects reviewed and the strongly related subjects are mentioned in appendix A and E respectively.
The results obtained through clustering gained important knowledge and insights that can be used for improving the performance of students. The yield was different clusters that is, cluster1: students attending the classes regularly scored high marks and cluster2: students attending regularly scored less marks. This result helps to predict whether scoring marks in a subject actually depends on attendance or not. It even helps to find out weak students in a particular subject. This will help the teachers to improve the performance of the students who are weak in those particular subjects. (see Appendix F).

TABLE III. INSTANCE OF A DATABASE FOR CLUSTERING

| Stud_id | attendance | marks |
|---------|------------|-------|
| 1 | 93 | 20 |
| 2 | 100 | 41 |
| 3 | 100 | 41 |
| 4 | 100 | 25 |
| 5 | 87 | 46 |
| . | . | . |
| . | . | . |

The result obtained from classification is a classifier in the form of decision tree which classifies the unseen student in order to predict the performance of the student. Prediction will help the teachers to pay attention to poor and average students in order to enhance their capabilities in their academics.
The result of Clustering and Classification is mentioned in appendix F and G respectively.

TABLE IV. INSTANCE OF A DATABASE FOR CLASSIFICATION

| Stud_id | Dept | Attendance | Marks | Performance |
|---------|------|------------|-------|-------------|
| 1 | ETC | Y | 310 | AVERAGE |
| 2 | IT | N | 450 | GOOD |
| 3 | COMP | Y | 500 | GOOD |
| 4 | IT | Y | 230 | POOR |
| . | . | . | . | . |
| . | . | . | . | . |

After applying classification algorithm we get a decision tree which is dependent on the "gain" (see Appendix G).

*B. Significance of Results*

The results obtained through our experiment gained important knowledge and insights that can be used for improving the quality of the educational programmes. Some of these insights are outlined below:

- Preconceived notion of a relationship between Mathematics subjects and programming subjects:

There existed a general notion that mathematics subjects and programming subjects are correlated. However, our experiments illustrated that there does not exists a strong relationship between these subjects. That is, passing or failing a mathematics subject does not determine the ability to pass/fail a programming subject and vice-versa.

- Assist in determining pre-requisite subjects:

When determining prerequisites it is advantageous to know that the existence of the strong relationship between subjects. A student may fail a particular subject (say subjectA) and proceed to taking further subjects (say subjects, subjectB, subjectC). However, the student may not have acquired the necessary knowledge and skills required (i.e. pre-requisite knowledge) for passing subjectB and subjectC. Hence, the student may fail with a high probability and waste student's and institution's resources. If there a large percentage of students who fail a pre-requisite subject (i.e. subjectA) also fail the subject (i.e. subjectB), then these subjects are strongly related (subjectA is strongly related to subjectB) and is captured in our experimental results.

- Project Course:

At the end of the 6th semester, RIEIT focuses on students completing a project as a team. The main objective of the course is to apply knowledge gained from other subjects to solve a real-world problem. So our experiment will be beneficial for the students to select project ideas which are based on present subject and related to past subject.

Appendix

*A. List of subject id and their titles*

| id | Subject |
|------|---------|
| IT31 | Applied Mathematics III |
| IT32 | Numerical Methods |
| IT33 | Analog And Digital Communication |
| IT34 | Computer Organization And Architecture |
| IT35 | Data Structures Using C |
| IT36 | System Analysis And Design |
| IT41 | Discrete Mathematical Structures |
| IT42 | Signals And Systems |
| IT43 | Computer Hardware And Troubleshooting |
| IT44 | Microprocessors And Interfaces |
| IT45 | Design And Analysis Of Algorithms |
| IT46 | Object Oriented Programming System |
| IT51 | Introduction To Data Communication |
| IT52 | Digital Signal Processing |
| IT53 | Software Engineering |
| IT54 | Intelligent Agents |
| IT55 | Operating Systems |

| IT56 | Database Management System |
| IT51 | Entrepreneurship Development |
| IT52 | Theory Of Computation |
| IT53 | Computer Networks |
| IT54 | Computer Graphics |
| IT55 | Web Technology |
| IT56 | Software Testing And Quality Assurance |
| ETC31 | Applied Mathematics III |
| ETC32 | Digital System Design |
| ETC33 | Network Analysis And Synthesis |
| ETC34 | Electronic Devices And Circuits |
| ETC35 | Managerial Economics |
| ETC36 | Computer Oriented Numerical Techniques |
| ETC41 | Applied Mathematics IV |
| ETC42 | Signals And Systems |
| ETC43 | Electrical Technology |
| ETC44 | Electromagnetic Field And Waves |
| ETC45 | Linear Integrated Circuits |
| ETC46 | Data Structures Using C++ |
| ETC51 | Probability Theory And Random Processes |
| ETC52 | Control System Engineering |
| ETC53 | Communication Engineering 1 |
| ETC54 | Microprocessors |
| ETC55 | Digital Signal Processing |
| ETC56 | Transmission Lines And Waveguides |
| ETC61 | Communication Engineering 2 |
| ETC62 | Peripheral Devices And Interfacing |
| ETC63 | Power Electronics |
| ETC64 | Antenna And Wave Propagation |
| ETC65 | Electronic Instrumentation |
| ETC66 | VLSI Technologies And Design |
| COMP31 | Applied Mathematics III |
| COMP32 | Basics Of C++ |
| COMP33 | Principles Of Programming Languages |
| COMP34 | Computer Oriented Numerical Techniques |
| COMP35 | Logic Design |
| COMP36 | Integrated Electronics |
| COMP41 | Discrete Mathematical Structures |
| COMP42 | Data Structures |
| COMP43 | Computer Organization |
| COMP44 | Electronic Measurements |
| COMP45 | System Analysis And Design |
| COMP46 | Object Oriented Programming & Design Using C++ |
| COMP51 | Organizational Behavior And Cyber Law |
| COMP52 | Automata Language And Computation |
| COMP53 | Microprocessors And Microcontrollers |
| COMP54 | Computer Hardware Design |
| COMP55 | Database Management System |
| COMP56 | Operating System |
| COMP61 | Modern Algorithm Design Foundation |
| COMP62 | Object Oriented Software Engineering |
| COMP63 | Artificial Intelligence |
| COMP64 | Computer Graphics |
| COMP65 | Device Interface And Pc Maintenance |
| COMP66 | Data Communications |

B. *Subjects offered in the IT stream for semesters 3-6*

| 3rd Semester | 4th Semester | 5th Semester | 6th Semester |
|---|---|---|---|
| IT31 | IT41 | IT51 | IT61 |
| IT32 | IT42 | IT52 | IT62 |
| IT33 | IT43 | IT53 | IT63 |
| IT34 | IT44 | IT54 | IT64 |
| IT35 | IT45 | IT55 | IT65 |
| IT36 | IT46 | IT56 | IT66 |

C. *Subjects offered in the ETC stream for semesters 3-6*

| 3rd Semester | 4th Semester | 5th Semester | 6th Semester |
|---|---|---|---|
| ETC31 | ETC41 | ETC51 | ETC61 |
| ETC32 | ETC42 | ETC52 | ETC62 |
| ETC33 | ETC43 | ETC53 | ETC63 |
| ETC34 | ETC44 | ETC54 | ETC64 |
| ETC35 | ETC45 | ETC55 | ETC65 |
| ETC36 | ETC46 | ETC56 | ETC66 |

D. *Subjects offered in the COMP stream for semesters 3-6*

| 3rd Semester | 4th Semester | 5th Semester | 6th Semester |
|---|---|---|---|
| COMP31 | COMP41 | COMP51 | COMP61 |
| COMP32 | COMP42 | COMP52 | COMP62 |
| COMP33 | COMP43 | COMP53 | COMP63 |
| COMP34 | COMP44 | COMP54 | COMP64 |
| COMP35 | COMP45 | COMP55 | COMP65 |
| COMP36 | COMP46 | COMP56 | COMP66 |

E. *Strongly Related subjects in the respective streams with $\gamma > 0.5$*

IT Stream

| | |
|---|---|
| ENTREPRENEURSHIP DEVELOPMENT | SOFTWARE TESTING AND QUALITY ASSURANCE |
| COMPUTER GRAPHICS | SOFTWARE TESTING AND QUALITY ASSURANCE |
| ENTREPRENEURSHIP DEVELOPMENT | COMPUTER GRAPHICS |
| DATA STRUCTURES USING C | DISCRETE MATHEMATICAL STRUCTURES |
| SOFTWARE ENGINEERING | SOFTWARE TESTING AND QUALITY ASSURANCE |
| DATA STRUCTURES USING C | DESIGN AND ANALYSIS OF ALGORITHMS |
| COMPUTER HARDWARE AND TROUBLESHOOTING | SOFTWARE TESTING AND QUALITY ASSURANCE |
| COMPUTER HARDWARE AND TROUBLESHOOTING | ENTREPRENEURSHIP DEVELOPMENT |
| SOFTWARE ENGINEERING | ENTREPRENEURSHIP DEVELOPMENT |
| SOFTWARE ENGINEERING | COMPUTER GRAPHICS |
| DISCRETE MATHEMATICAL STRUCTURES | ENTREPRENEURSHIP DEVELOPMENT |
| COMPUTER HARDWARE AND TROUBLESHOOTING | COMPUTER GRAPHICS |
| DESIGN AND ANALYSIS OF ALGORITHMS | ENTREPRENEURSHIP DEVELOPMENT |
| COMPUTER HARDWARE AND TROUBLESHOOTING | DESIGN AND ANALYSIS OF ALGORITHMS |
| DATA STRUCTURES USING C | SOFTWARE TESTING AND QUALITY ASSURANCE |
| INTRODUCTION TO DATA COMMUNICATION | COMPUTER GRAPHICS |
| DATA STRUCTURES USING C | ENTREPRENEURSHIP DEVELOPMENT |
| DISCRETE MATHEMATICAL STRUCTURES | DESIGN AND ANALYSIS OF ALGORITHMS |

COMP Stream

| | |
|---|---|
| DEVICE INTERFACE AND PC MAINTENANCE | DATA COMMUNICATIONS |
| OPERATING SYSTEM | DEVICE INTERFACE AND PC MAINTENANCE |
| BASICS OF C++ | DATA STRUCTURES |
| BASICS OF C++ | DEVICE INTERFACE AND PC MAINTENANCE |
| ORGANISATIONAL BEHAVIOUR AND CYBER LAW | DEVICE INTERFACE AND PC MAINTENANCE |
| COMPUTER ORGANISATION | DEVICE INTERFACE AND PC MAINTENANCE |
| SYSTEM ANALYSIS AND DESIGN | DEVICE INTERFACE AND PC MAINTENANCE |
| ORGANISATIONAL BEHAVIOUR AND CYBER LAW | DATA COMMUNICATIONS |
| DATA STRUCTURES | DATA COMMUNICATIONS |
| COMPUTER ORGANISATION | DATA COMMUNICATIONS |
| OBJECT ORIENTED PROGRAMMING AND DESIGN USING C++ | DEVICE INTERFACE AND PC MAINTENANCE |
| BASICS OF C++ | COMPUTER ORGANISATION |
| ELECTRONIC MEASUREMENTS | DEVICE INTERFACE AND PC MAINTENANCE |
| COMPUTER ORGANISATION | ELECTRONIC MEASUREMENTS |
| OBJECT ORIENTED SOFTWARE ENGINEERING | COMPUTER GRAPHICS |
| COMPUTER GRAPHICS | DATA COMMUNICATIONS |
| BASICS OF C++ | DATA COMMUNICATIONS |
| BASICS OF C++ | SYSTEM ANALYSIS AND DESIGN |

ETC Stream

| | |
|---|---|
| ELECTRONIC INSTRUMENTATION | VLSI TECHNOLOGIES AND DESIGN |
| NETWORK ANALYSIS AND SYNTHESIS | SIGNALS AND SYSTEMS |
| ELECTRONIC DEVICES AND CIRCUITS | SIGNALS AND SYSTEMS |
| NETWORK ANALYSIS AND SYNTHESIS | COMPUTER ORIENTED NUMERICAL TECHNIQUES |
| ELECTRONIC DEVICES AND CIRCUITS | ELECTROMAGNETIC FIELD AND WAVES |
| ELECTRONIC DEVICES AND CIRCUITS | MANAGERIAL ECONOMICS |
| MANAGERIAL ECONOMICS | COMPUTER ORIENTED NUMERICAL TECHNIQUES |
| ELECTROMAGNETIC FIELD AND WAVES | LINEAR INTEGRATED CIRCUITS |
| DIGITAL SIGNAL PROCESSING | VLSI TECHNOLOGIES AND DESIGN |
| ELECTRONIC DEVICES AND CIRCUITS | APPLIED MATHEMATICS IV |
| ELECTRONIC DEVICES AND CIRCUITS | DATA STRUCTURES USING C++ |
| COMPUTER ORIENTED NUMERICAL TECHNIQUES | SIGNALS AND SYSTEMS |
| MANAGERIAL ECONOMICS | SIGNALS AND SYSTEMS |
| NETWORK ANALYSIS AND SYNTHESIS | ELECTRONIC DEVICES AND CIRCUITS |
| ELECTRONIC DEVICES AND CIRCUITS | ELECTRICAL TECHNOLOGY |
| SIGNALS AND SYSTEMS | ELECTRICAL TECHNOLOGY |
| SIGNALS AND SYSTEMS | ELECTROMAGNETIC FIELD AND WAVES |
| NETWORK ANALYSIS AND SYNTHESIS | DATA STRUCTURES USING C++ |

F. *Clusters obtained in the respective streams*

IT Stream

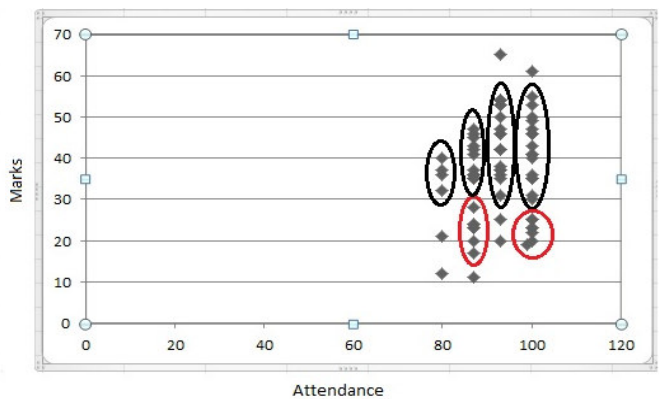

Computer Engg. Stream

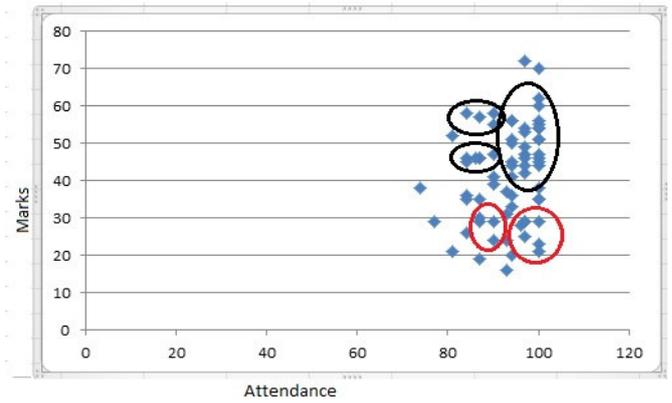

ETC Stream

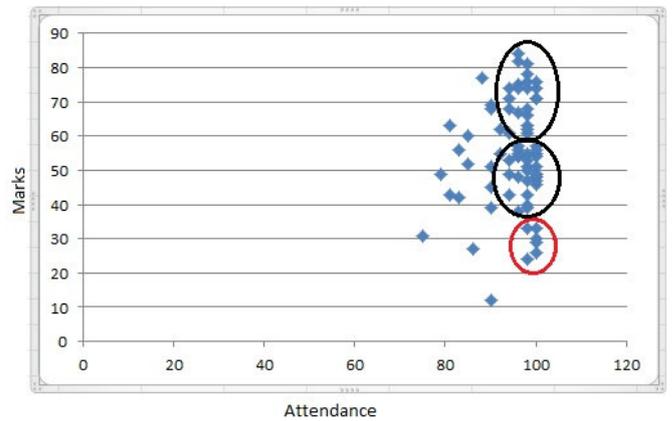

G. *Decision Tree obtained as a result of classification.*

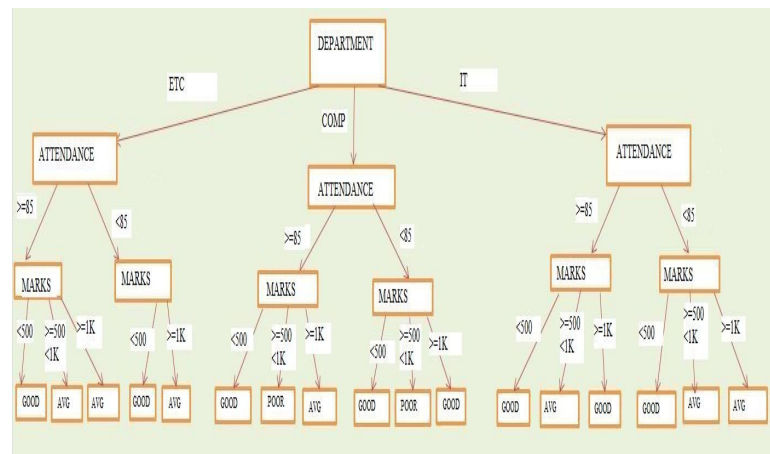